\documentclass[nofootinbib, %
 reprint,
 amsmath,amssymb,
 aps, showkeys
]{revtex4-1}

\usepackage{graphicx}% Include figure files
\usepackage{dcolumn}% Align table columns on decimal point
\usepackage{bm}% bold math
\usepackage{float}
\begin{document}

%\preprint{APS/123-QED}

\title{Finding Exact Forms on a Thermodynamic Manifold}% Force line breaks with \\
%\thanks{A footnote to the article title}%

\author{Chao Ju}
 \affiliation{Minerva Schools at KGI}%Lines break automatically or can be forced with \\
\author{Mark Stalzer}%
 %\email{stalzer@caltech.edu}
\affiliation{
 California Institute of Technology\\
}

%\collaboration{MUSO Collaboration}%\noaffiliation

%\author{Charlie Author}
% \homepage{http://www.Second.institution.edu/~Charlie.Author}
%\affiliation{
% Second institution and/or address\\
% This line break forced% with \\
%}
%\affiliation{
% Third institution, the second for Charlie Author
%}
%\author{Delta Author}

%\collaboration{CLEO Collaboration}%\noaffiliation

\date{\today}% It is always \today, today,
             %  but any date may be explicitly specified

\begin{abstract}
Because only two variables are needed to characterize a simple thermodynamic system in equilibrium, any such system is constrained on a 2D manifold. Of particular interest are the exact 1-forms on the cotangent space of that manifold, since the integral of exact 1-forms is path-independent, a crucial property satisfied by state variables such as the internal energy $dE$ and the entropy $dS$. Our prior work\cite{stalzer} shows that given an appropriate \textit{language} of vector calculus, a machine can re-discover the Maxwell equations and the incompressible Navier-Stokes equations from simulation. We speculate that we can enhance this language by including differential forms. In this paper, we use the example of classical thermodynamics to show that there exists a simple algorithm to automate the process of finding exact 1-forms on a thermodynamic manifold. Since entropy appears in various fields of science in different guises, a potential extension of this work is to use the machinery developed in this paper to re-discover the expressions for entropy from data in fields other than classical thermodynamics.
%\begin{description}
%\item[Usage]
%Secondary publications and information retrieval purposes.
%\item[PACS numbers]
%May be entered using the \verb+\pacs{#1}+ command.
%\item[Structure]
%You may use the \texttt{description} environment to structure your abstract;
%use the optional argument of the \verb+\item+ command to give the %category of each item. 
%\end{description}
 
\end{abstract}
\pacs{Valid PACS appear here}% PACS, the Physics and Astronomy
                             % Classification Scheme.
%\keywords{Suggested keywords}%Use showkeys class option if keyword
                              %display desired
                        
\maketitle

%\tableofcontents

\section{Introduction}
We model the system of interest using an ideal gas of a certain volume. The system is allowed to contract and expand, exchange heat with the surroundings, and do work, assuming the processes are quasi-static. We can of course represent the state of the system on a p-V diagram, but that hides much of the richness of the system. If, instead, we treat the system as a submanifold of $\mathbb{R}^3$, we will discover much structure by using the language of exterior calculus, one that deals with differential forms on manifolds\cite{Frankel}. \par
\begin{figure}[H]
\centering
\includegraphics[scale=0.4]{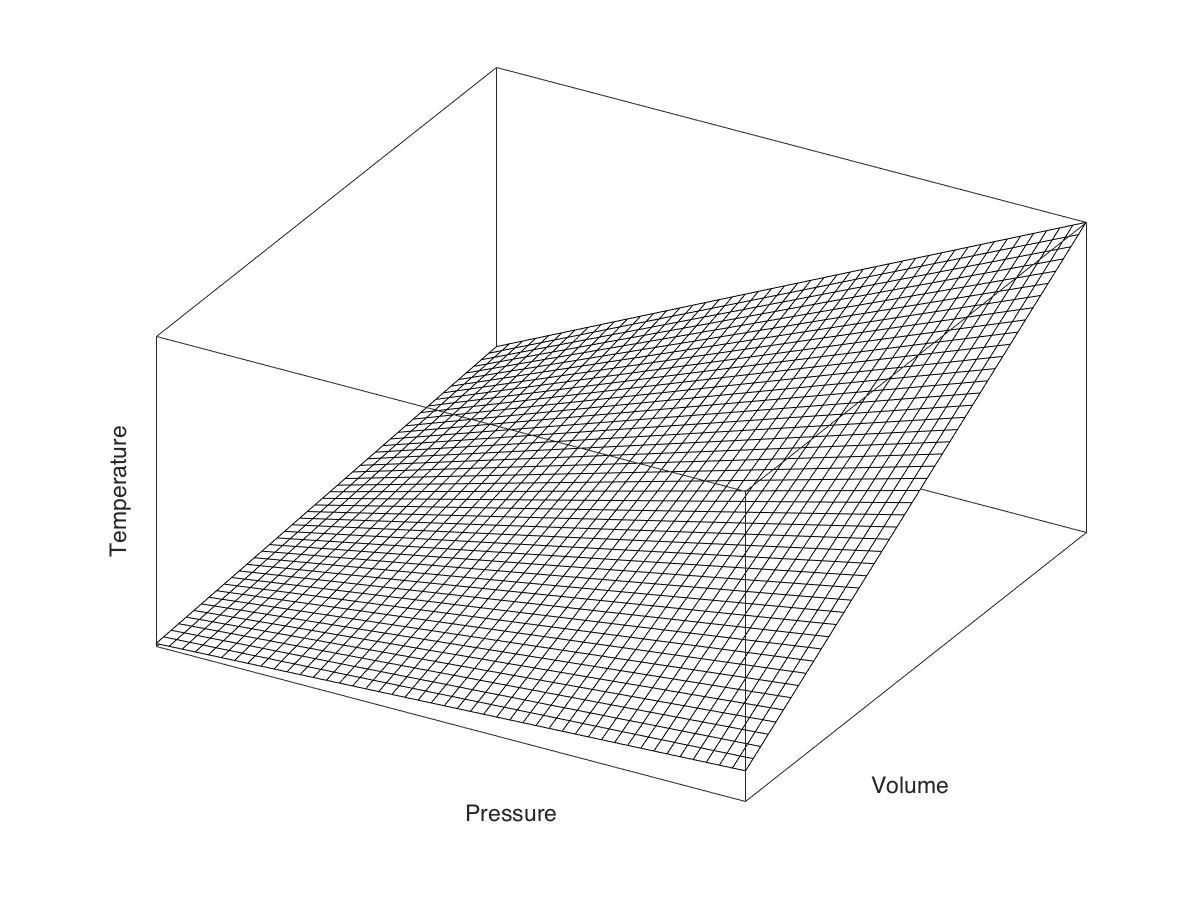}
\caption{The mesh shows a part of the 2D thermodynamic submanifold embedded in $\mathbb{R}^3$ for an ideal gas $pV=Nk_BT$. The state of the gas is represented by a point on the submanifold.}
\end{figure}
Let that submanifold occupied by the simple thermodynamic system be denoted by $M^2$, the superscript indicating that the submanifold is locally $\mathbb{R}^2$. Then, the space of 1-forms at a point on $M^2$ has dimension
\[
\dim \bigwedge^1(\mathbb{R}^2) = \binom{2}{1}=2
\]\par
Therefore, on this submanifold, we can expand any 1-form in any basis consisting of the differentials of a set of 2 coordinate functions, $dx$ and $dy$. For example, in Caratheodory's formulation of thermodynamics\cite{Frankel}, the first law reads
\begin{equation}
dE = Q^1 - W^1
\end{equation}
where $Q^1$ is the heat 1-form and $W^1$ is the work 1-form. The superscript refers to the dimension of the form. Given the knowledge of entropy and work, we can expand the above equation as
\begin{align*}
dE &= TdS - pdV \\
&= T(S,V) dS - p(S,V) dV
\end{align*} \par
The above example shows that indeed, $dE$ as a 1-form can be expanded in a basis $\{dS,dV\}$ using only the variables $S$ and $V$. In the following analysis, without loss of generality\footnote{Indeed, forms are geometric objects whose properties are coordinate-independent. Exact forms in one basis stay exact in any other basis.} we pick the basis of our 1-forms as $\{dp, dV\}$ with the goal of expressing every 1-form in terms of $p$ and $V$. In addition to the variables $p$ and $V$, we have 2 other important ``constants'', $nR$ and $c_v$, where the first combination comes from the ideal gas law $pV=nRT$, and $c_v$ is the heat capacity which appears in $dE=c_vdT$.
\section{A general expression for exact 1-forms}
By definition, the goal is to look for any 1-form $f^1$ such that $f^1=dg$, where $g=g(p,V)$ is a 0-form function. Instead of enumerating all possible $g$ and taking the differential to get $f^1$, we observe that on the thermodynamic manifold, all closed 1-forms are exact. This is because of the De Rham cohomology of this manifold, which has vanishing first Betti number $b_1=0$. By De Rham's theorem, all closed 1-forms on this manifold are therefore exact. Since all exact 1-forms are automatically closed, to find those exact forms we can simply look for closed 1-forms $f^1$ such that $df^1=0$. As we shall see, the condition $df^1=0$ severely constrains the form $f$ can take, and will reduce the enumeration space enormously.

\par
Using the $\{dp, dV\}$ basis, we can express every 1-form $f^1$ on the thermodyanmic submanifold as
\begin{equation}
f^1 = A(p,V) dp  + B(p,V) dV
\end{equation}
where $A$ and $B$ are assembled from the symbols of the following set $\mathcal{S}$. Note that consistent with our previous approach\cite{stalzer}, we exclude any transcendental functions in the language.
\[
\mathcal{S} = \{p, V, nR, c_v, +, -, \times, \div \}
\] \par
At first sight, enumerating all possible $f^1$ seems a daunting task, because the enumeration space is too big. However, there are two crucial pieces of information that we can harness to significantly reduce the size of the enumeration space.\par
\textit{First}, we demand that the units of the 2 summands in equation (2) must agree. This constraint is a physical one that must be satisfied by any equation. The first consequence of this constraint is that we can leave $Nk_B$ and $c_v$ out of the enumeration space for a while: both of their units contain $1/[Temperature]$, which does not appear in the unit of $p$ or $V$. Therefore, they must have the same power in $A$ and $B$ to balance out the temperature. \par
The second consequence of this constraint on unit is that, if we write
\begin{equation}
f^1 = p^\alpha V^\beta dp + p^{\alpha'} V^{\beta'} dV
\end{equation}
then we will obtain 2 independent linear equations
\begin{align}
\alpha' &= \alpha + 1 \\
-\alpha + 3\beta - 1 & = -\alpha' + 3\beta' +3
\end{align}
by dimensional analysis. We have now used up the information of the first constraint.\par
The second constraint on $f^1$ is closedness: the goal is to enumerate closed 1-forms only. That said, we want $f^1$ such that 
\[
df^1 = 0
\]
which, from (1), is
\begin{align*}
%dA \wedge dp + dB \wedge dV &= 0 \Rightarrow \\
%\frac{\partial A}{\partial V} dV \wedge dp +  \frac{\partial %B}{\partial p} dp \wedge dV &= 0 \Rightarrow \\
%(\frac{\partial A}{\partial V}-\frac{\partial B}{\partial p}) dV \wedge dp &= 0 \Rightarrow \\ 
\frac{\partial A}{\partial V}  &= \frac{\partial B}{\partial p}
\end{align*}
Therefore, if in (3) we assume that $\beta\neq 0$ and $\alpha' \neq 0$, then by equating the partial derivatives we obtain another linear equation
\begin{equation}
-\alpha + 3\beta - 4 = -\alpha' + 3\beta' + 4
\end{equation}
but this leads to a contradiction with (5). Therefore, we must have 
\[
\beta= \alpha'=0
\]
and this combined with (5) gives us
\[
\alpha = \beta' = -1
\] \par
Therefore, (3) becomes 
\[
f^1 = \frac{1}{p} dp + \frac{1}{V} dV
\]
and if we merge the previously left-out $Nk_B$ and $c_v$ into the constants $c_1$ and $c_2$, we obtain the final ansatz of our closed 1-form:
\begin{equation}
f^1 = \frac{c_1}{p}dp + \frac{c_2}{V} dV
\end{equation}
where $c_1$ and $c_2$ are constants of the same dimension.

\section{Entropy}
A valid thermodynamic theorem (equation) must equate n-forms to n-forms. The first law, equation (1), is one such example that equates 1-forms to 1-forms. This section is concerned with finding a thermodynamic theorem governing entropy for a simple system in equilibrium. \par

In our prior work on Maxwell and Navier-Stokes, we created a program to enumerate ``theorems'' (instantiated by equations) from a set of symbols, and then validate a certain theorem by using the output of a virtual experiment to see whether the constants in the theorem can be found. To start with, we need to create a finite set consisting of singleton theorems
\[
\mathcal{H}=\{A_1, A_2, A_3, ...\}
\]
where each singleton theorem $A_i$ is associated with a certain complexity score and is represented by a linear equation
\[
c_0 + c_1 A_i = 0
\]
where $c_0$ and $c_1$ are constants to be found by the program to test the validity of the theorem. A concrete example for a singleton theorem is when $A_1=\nabla\cdot\textbf{B}$, where $\textbf{B}$ is the magnetic field. Then the first singleton theorem enumerated from the set is 
\[
c_0 + c_1 \nabla\cdot\textbf{B}  = 0
\]
and electromagnetism tells us that this is a valid theorem for $c_0=0$ and $c_1=1$. \par 
After we input the singleton theorem set, the program takes another input $N$, the total complexity score, and efficiently enumerates all candidate theorems whose complexity scores are no more than $N$\cite{stalzer2}. For example, suppose each $A_i$ in the set $\mathcal{H}$ has a complexity score of 1, then theorems of complexity score $2$ are of the following form: 
\[
c_0 + c_1 A_i + c_2 A_j = 0, \quad\forall i\neq j
\]\par 
The program uses a smart way to validate a theorem as soon as it is enumerated by using the output of a virtual experiment. For example, the virtual experiment we used to re-discover the Maxwell equations is the far-field behavior of an oscillating electric dipole with a certain angular frequency and dipole moment\cite{stalzer}. From the output of this virtual experiment the program can validate theorems such as $c_0 + c_1 \nabla\cdot\textbf{B}  = 0$. The method of validating theorems involves the use of applied linear algebra, and the details can be found in \cite{stalzer}.\par
The above is a summary of the essential process of enumerating and validating theorems. In the following, we shall show that an expression for entropy can be found using this process. \par 
To start with, we hypothesize that entropy $S$ is an observable of a certain virtual experiment\footnote{The assumption of entropy as an observable might be a bit far-fetched. However, just as work (which itself is not a direct observable) can be obtained by measuring force and distance, so entropy can be obtained by calculating heat and measuring temperature. The purpose here is to show how the process of finding theorems works.}, and that its differential $dS$ is a 1-form. Then, using the theoretical results obtained from the previous section, we can form a tentative theorem set
\[
\mathcal{T}=\{dS, \frac{1}{V}dV, \frac{1}{p}dp\}
\]\par 
One theorem that is guaranteed to be enumerated from $\mathcal{T}$ is 
\begin{equation}
    c_0 + c_1 dS+ c_2\frac{1}{p}dp+ c_3 \frac{1}{V}dV=0
\end{equation} \par 
To test whether the above equation is a valid theorem or not, we must use the output from a certain virtual experiment and solve a system of linear equations to find the constants. If the constants have a unique nontrivial solution, then we conclude that (8) is a valid theorem. In this application, we shall simply use 1 mole of monatomic gas that can contract and expand as the virtual experiment, whose output for entropy has a simple mathematical expression valid for moderate temperature\cite{fermi}
\begin{equation}
S=c_v \ln{\frac{pV}{R}} + R \ln{V} + a
\end{equation}
where $a$ is a constant whose specific value is irrelevant in this application: to set up equations, we want the difference in entropy instead of its absolute value. In general, the virtual experiment can be represented by a trajectory $x(t)$ on the p-V diagram parameterized by $t$:
\[
x(t)=(p(t), V(t))
\]
and the output of the virtual experiment is $S(t)=S(p(t),V(t))$. To validate the theorem, we need to pull back (8) onto the $t$ variable and evaluate the integral
\begin{align*}
c_1\Delta S &= c_2'\int F_t^*\{\frac{1}{p}dp\}+c_3'\int F_t^*\{\frac{1}{V}dV\}  \\
&=c_2'\int_{t_1}^{t_2} \frac{1}{p(t)}\frac{dp}{dt}dt + c_3'\int_{t_1}^{t_2}\frac{1}{V(t)}\frac{dV}{dt}dt
\end{align*}
where $\Delta S=S(t_2)-S(t_1)$, $F^*_t$ is the pull-back from the p-V plane to $t$, and $c_2'=-c_2$, $c_3'=-c_3$. We can then merge $c_1$ into the other 2 constants to obtain the following equation: 
\begin{equation}
\Delta S=c_2'\int_{t_1}^{t_2} \frac{1}{p(t)}\frac{dp}{dt}dt + c_3'\int_{t_1}^{t_2}\frac{1}{V(t)}\frac{dV}{dt}dt
\end{equation} \par 
In most applications, the output data of the virtual experiment come in discrete forms: 
\[
\{p(t_i), V(t_i), S(t_i)\}
\]
and we need to numerically integrate (10) and set up equations to find $c_2'$ and $c_3'$ given a trajectory. In the following, we use a simplified trajectory to finalize this example with the goal of showing the essentials while avoiding numerical integrations.

\par To turn (10) into a set of linear equations, we specify 3 points $A=(p_1, V_1)$, $B=(p_2, V_1)$, $C=(p_2,V_2)$. Starting at point $A$, we integrate (8) isochorically to point $B$, and then isobarically to point $C$.
\begin{figure}[H]
\centering
\includegraphics[scale=0.4]{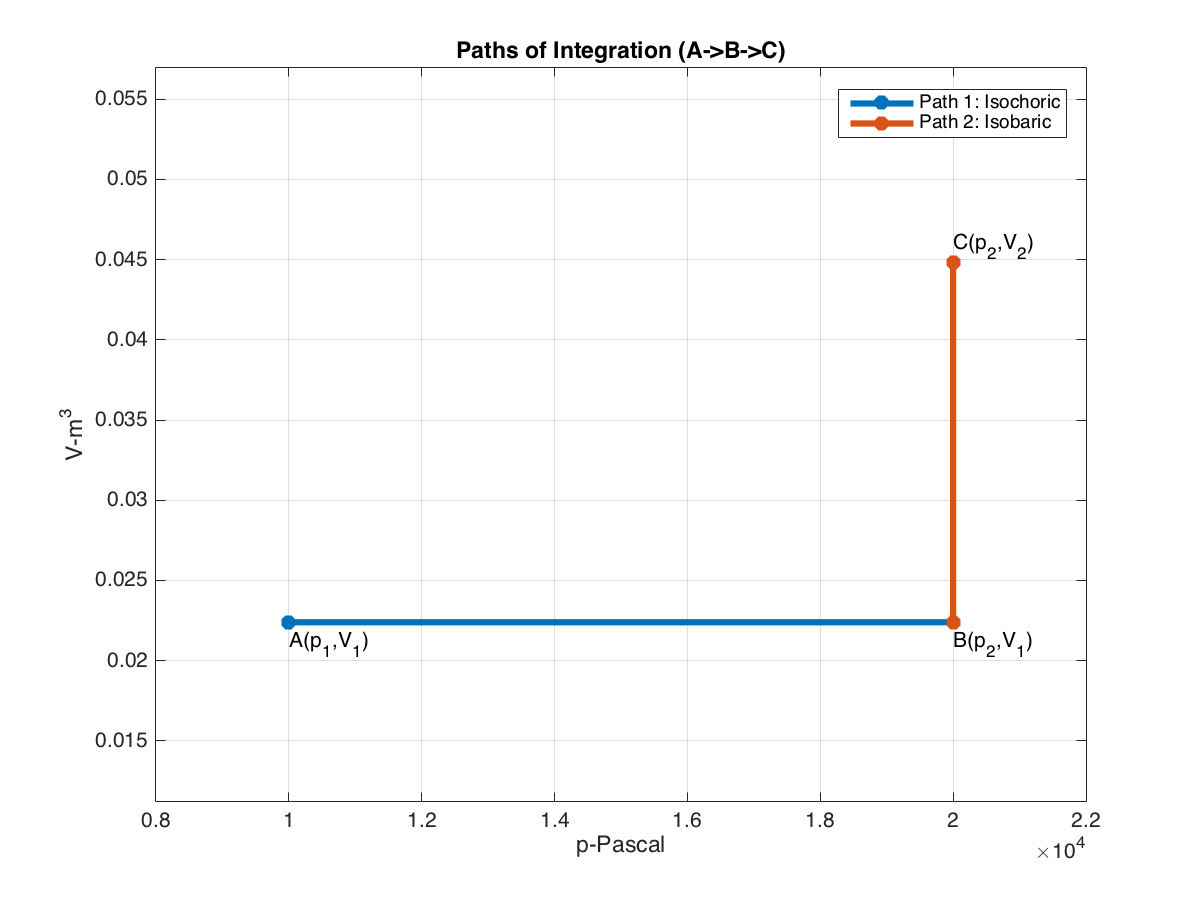}
\caption{Path of integration from $A$ to $B$ to $C$.}
\end{figure}\par
The 2 equations we obtain are thus
\begin{align}
    S(B)-S(A) &= \int_{p_1}^{p_2} \frac{c_2'}{p}dp = c_2' \ln{\frac{p_2}{p_1}} \\
    S(C)-S(B) &= \int_{V_1}^{V_2} \frac{c_3'}{V}dV = c_3' \ln{\frac{V_2}{V_1}}
\end{align}\par
Let $p_1=10000$ Pa, $V_1=22.4\times 10^{-3}$ m$^3$ (this is the approximate volume of 1 mole of ideal gas at standard room temperature and pressure), and $p_2=2p_1$, $V_2=2V_1$. The virtual experiment (instantiated by (9)) gives us the following output (given $R=8.3145$ J/(mol K) and $c_v=3/2R$):
\begin{align*}
    S(B)-S(A) &= 8.644758 J/K \\
    S(C)-S(B) &= 14.407931 J/K
\end{align*} \par
From the above output of the virtual experiments, we can then solve for $c_2'$ and $c_3'$ in equations (11) and (12). They are
\begin{align*}
    c_2' &= 12.47175 J/K \\
    c_3' &= 20.78625 J/K
\end{align*}
and we conclude that (8) is a valid theorem. In fact, given the knowledge of thermodynamics, we can easily show that $c_2'=c_v$, $c_3'=c_v+R$, and the correct expression for $dS$ for 1 mole of ideal gas is
\[
dS = \frac{c_v}{p} dp + \frac{c_v+R}{V} dV + c_0
\] 
where $c_0$ is an additive constant. In this example, we used a simplified approach to illustrate the core idea of constructing tentative theorems from a given set and the use of virtual experiment to determine the validity of a theorem. The complete process can be found in \cite{stalzer}.

\section{Conclusion}
We have shown that we can greatly simplify the problem of enumerating exact 1-forms using one mathematical (closedness) and one physical (dimensional analysis) constraint. In our previous work, we dealt with re-discovering linear differential theories using the language of vector calculus. The above result shows that there is great potential to extend our previous framework to cover exterior calculus, which will enable us in the future to re-discover scientific theorems that can be geometrically formulated. As an example, Lott and Villani\cite{villani} have established a mathematical connection between Ricci curvature, entropy, and optimal transport. But Ricci curvature, $R_{\mu\nu}$, can be thought of as a vector-valued 1-form when the first index is raised by some metric $R^\mu_{\ \nu}=g^{\mu\sigma}R_{\sigma\nu}$. In addition, another vector-valued measure of curvature is $\theta^{\mu}_{\ \nu} = \frac{1}{2} R^{\mu}_{\ \nu\rho\sigma} dx^\rho \wedge dx^\sigma$. Perhaps, given a judicious choice of singleton theorem set and virtual experiment, we could find some curious functional relationship between curvature, entropy (which might also appear as a vector-valued 1-form by covariance), and other physical variables in the transport setting or gradient flow.
\par
In re-discovering old theorems, we wish to establish the robustness of this enumeration-validation framework, but the ultimate goal is to apply this framework to find new scientific laws from a wealth of data available that could shed light on scientific discovery.


\begin{thebibliography}{9}
\bibitem{stalzer} M. Stalzer, C. Ju.
\textit{TheoSea: Marching Theory to Light}.
arXiv:1708.04927 [cs.AI]. This work expanded with an application to Navier-Stokes equations has been submitted to \textit{Artificial Intelligence}.

\bibitem{Frankel} T. Frankel.
\textit{The Geometry of Physics: An Introduction, 3rd Edition}.
Cambridge University Press, 2011.

\bibitem{fermi} E. Fermi.
\textit{Thermodynamics}.
Dover Publications, 2012.

\bibitem{stalzer2} M. Stalzer.
\textit{On the Enumeration of Sentences By Compactness}.
arXiv:1706.06975 [cs.AI].

\bibitem{villani} J. Lott, C. Villani.
\textit{Ricci Curvature for Metric-Measure Spaces via Optimal Transport}.
arXiv:math/0412127 [math.DG].

\end{thebibliography}
\end{document}